\documentclass[showpacs,aps,twocolumn,superscriptaddress,prb,amsmath]
{revtex4}
\usepackage{txfonts}
\usepackage{color}
\usepackage[latin9]{inputenc}
\usepackage{amssymb}
\usepackage{graphicx}
\usepackage{dcolumn}
\usepackage{bm}
\usepackage{hyperref}
\usepackage{color}

\begin{document}
\title{Theoretical prediction of a low-energy Stone-Wales graphene with intrinsic type-III Dirac-cone}
\author{Zhenhao Gong$^\sharp$}
\affiliation{Hunan Key Laboratory of Micro-Nano Energy Materials and Devices, Xiangtan University, Hunan 411105, P. R. China}
\affiliation{Laboratory for Quantum Engineering and Micro-Nano Energy Technology and School of Physics and Optoelectronics, Xiangtan University, Hunan 411105, P. R. China}
\author{Xizhi Shi$^\sharp$}
\affiliation{Hunan Key Laboratory of Micro-Nano Energy Materials and Devices, Xiangtan University, Hunan 411105, P. R. China}
\affiliation{Laboratory for Quantum Engineering and Micro-Nano Energy Technology and School of Physics and Optoelectronics, Xiangtan University, Hunan 411105, P. R. China}
\author{Jin Li}
\email{lijin@xtu.edu.cn}
\affiliation{Hunan Key Laboratory of Micro-Nano Energy Materials and Devices, Xiangtan University, Hunan 411105, P. R. China}
\affiliation{Laboratory for Quantum Engineering and Micro-Nano Energy Technology and School of Physics and Optoelectronics, Xiangtan University, Hunan 411105, P. R. China}
\author{Shifang Li}
\affiliation{Hunan Key Laboratory of Micro-Nano Energy Materials and Devices, Xiangtan University, Hunan 411105, P. R. China}
\affiliation{Laboratory for Quantum Engineering and Micro-Nano Energy Technology and School of Physics and Optoelectronics, Xiangtan University, Hunan 411105, P. R. China}
\author{Chaoyu He}
\email{hechaoyu@xtu.edu.cn}
\affiliation{Hunan Key Laboratory of Micro-Nano Energy Materials and Devices, Xiangtan University, Hunan 411105, P. R. China}
\affiliation{Laboratory for Quantum Engineering and Micro-Nano Energy Technology and School of Physics and Optoelectronics, Xiangtan University, Hunan 411105, P. R. China}
\author{Tao Ouyang}
\affiliation{Hunan Key Laboratory of Micro-Nano Energy Materials and Devices, Xiangtan University, Hunan 411105, P. R. China}
\affiliation{Laboratory for Quantum Engineering and Micro-Nano Energy Technology and School of Physics and Optoelectronics, Xiangtan University, Hunan 411105, P. R. China}
\author{Chunxiao Zhang  }
\affiliation{Hunan Key Laboratory of Micro-Nano Energy Materials and Devices, Xiangtan University, Hunan 411105, P. R. China}
\affiliation{Laboratory for Quantum Engineering and Micro-Nano Energy Technology and School of Physics and Optoelectronics, Xiangtan University, Hunan 411105, P. R. China}
\author{Chao Tang}
\affiliation{Hunan Key Laboratory of Micro-Nano Energy Materials and Devices, Xiangtan University, Hunan 411105, P. R. China}
\affiliation{Laboratory for Quantum Engineering and Micro-Nano Energy Technology and School of Physics and Optoelectronics, Xiangtan University, Hunan 411105, P. R. China}
\author{Jianxin Zhong}
\affiliation{Hunan Key Laboratory of Micro-Nano Energy Materials and Devices, Xiangtan University, Hunan 411105, P. R. China}
\affiliation{Laboratory for Quantum Engineering and Micro-Nano Energy Technology and School of Physics and Optoelectronics, Xiangtan University, Hunan 411105, P. R. China}

\begin{abstract}
Based on first-principles method we predict a new low-energy Stone-Wales graphene SW40, which has an orthorhombic lattice with Pbam symmetry and 40 carbon atoms in its crystalline cell forming well-arranged Stone-Wales patterns. The calculated total energy of SW40 is just about 133 meV higher than that of graphene, indicating its excellent stability exceeds all the previously proposed graphene allotropes. We find that SW40 processes intrinsic Type-III Dirac-cone (Phys. Rev. Lett., 120, 237403, 2018) formed by band-crossing of a local linear-band and a local flat-band, which can result in highly anisotropic Fermions in the system. Interestingly, such intrinsic type-III Dirac-cone can be effectively tuned by inner-layer strains and it will be transferred into Type-II and Type-I Dirac-cones under tensile and compressed strains, respectively. Finally, a general tight-binding model was constructed to understand the electronic properties nearby the Fermi-level in SW40. The results show that type-III Dirac-cone feature can be well understood by the $\pi$-electron interactions between adjacent Stone-Wales defects.
\end{abstract}

\maketitle
\indent The experimental synthesizing of two-dimensional (2D) graphene \cite{yhc1, yhc10, yhc11} and graphdiynes \cite{yhc2,yhc14} opened the door to the 2D carbon-word and have attracted much scientific efforts to reveal their fundamental properties and potential applications \cite{yhc1,yhc5,yhc6,yhc7,yhc8}. With excellent mechanical and electronic properties \cite{yhc1, yhc5, yhc6}, graphene is believed as a potential candidate for replacing silicon in future nano-electronics as new building block \cite{kkk}. To design pure-carbon nano-divice, 2D carbon allotropes can provide us rich electronic properties for different functional requirements. For example, R$_{57-1}$ \cite{yhc15}, R$_{57-2}$\cite{yhc17}, H$_{567}$ \cite{yhc17}, O$_{567}$ \cite{yhc17}, $\psi$-graphene \cite{yhc18, yhc19}, OPG-L \cite{prb2013}, net-$\tau$ \cite{carbon2019} and other 2D carbon allotropes \cite{yhc20, yhc21, yhc22, yhc23, yhc26, yhc27, yhc28, yhcprb, CMSncy2019} with normal metallic property can be used as electron conductors. The semiconducting octite SC \cite{yhc23}, pza-C10 \cite{yhc31}, $\Theta$-graphene \cite{YWC2018, ESM2019} and $\gamma$-graphyne \cite{prb2012, njp2013} are proper candidates \cite{yhcprb, NPGcm2019, NPGcm2018} for building diodes and transistors. The graphene\cite{yhc1, yhc10, yhc11}, phagraphene \cite{yhc36}, OPG-Z \cite{prb2013} and SW-graphene \cite{yhcprb} as Dirac-cone semi-metals \cite{yhcprb, NPGcm2019, NPGcm2018} with high carrier mobility can be used to construct high-speed nano-device. Especially, the freedom of rotation in graphene bilayer bring us the surprising phenomenon of superconductivity in some magic degrees \cite{magic1e, magic2e, magic3e, magic4e}, which has set off a new round of research upsurge on low-dimensional carbon systems \cite{magic1t, magic2t, magic3t, CYPprb}.

Recently, Dirac-cones with highly anisotropic property have attracted tremendous attentions as the great importance for the direction-dependent optical and electronic properties. It is reported that the anisotropic band dispersions can be induced in graphene through external Periodic Potentials\cite{NP08,PRL08} or elemental doping \cite{Gdoping}. In fact, intrinsic anisotropic Dirac-cones can be realized through graphene allotropes \cite{prb2013, yhc36, yhcprb} and graphyne allotropes \cite{O17, JPCL15, CYPprb, NPGcm2018}, such as OPG-Z\cite{prb2013}, phagraphene \cite{yhc36} and SW-graphene\cite{yhcprb}. However, the anisotropies of the Dirac-cones are usually not very strong. Recently, new types of Dirac cones, i.e.type-II\cite{PRB2016II,PRX2016,PRB2017,NP} and type-III Dirac cones\cite{PRL2018Liu}, have been proposed and attracted tremendous attentions. These new types of Dirac fermions have different properties with type-I Dirac fermions \cite{PRB2016II}, especially with very high anisotropic Fermi velocities around the Dirac cone. However, the type-II and type-III Dirac fermions seem to be much less common the type-I Dirac fermions in pure 2D carbon systems. We notice that only OPG-Z \cite{prb2013} possesses remarkable anisotropic properties due to its Type-III like Dirac-cone \cite{PRL2018Liu}, however, its calculated total energy is 345 meV/atom higher than that of graphene, indicating that it is difficult to be synthesized in future experiments. Therefore, to theoretically propose experimentally viable 2D carbons with type-II or type-III Dirac fermions \cite{PRL2018Liu} and highly anisotropic properties is of crucial importance for both fundamental and practical interests.
\begin{figure*}
\begin{center}
\includegraphics[width=\textwidth]{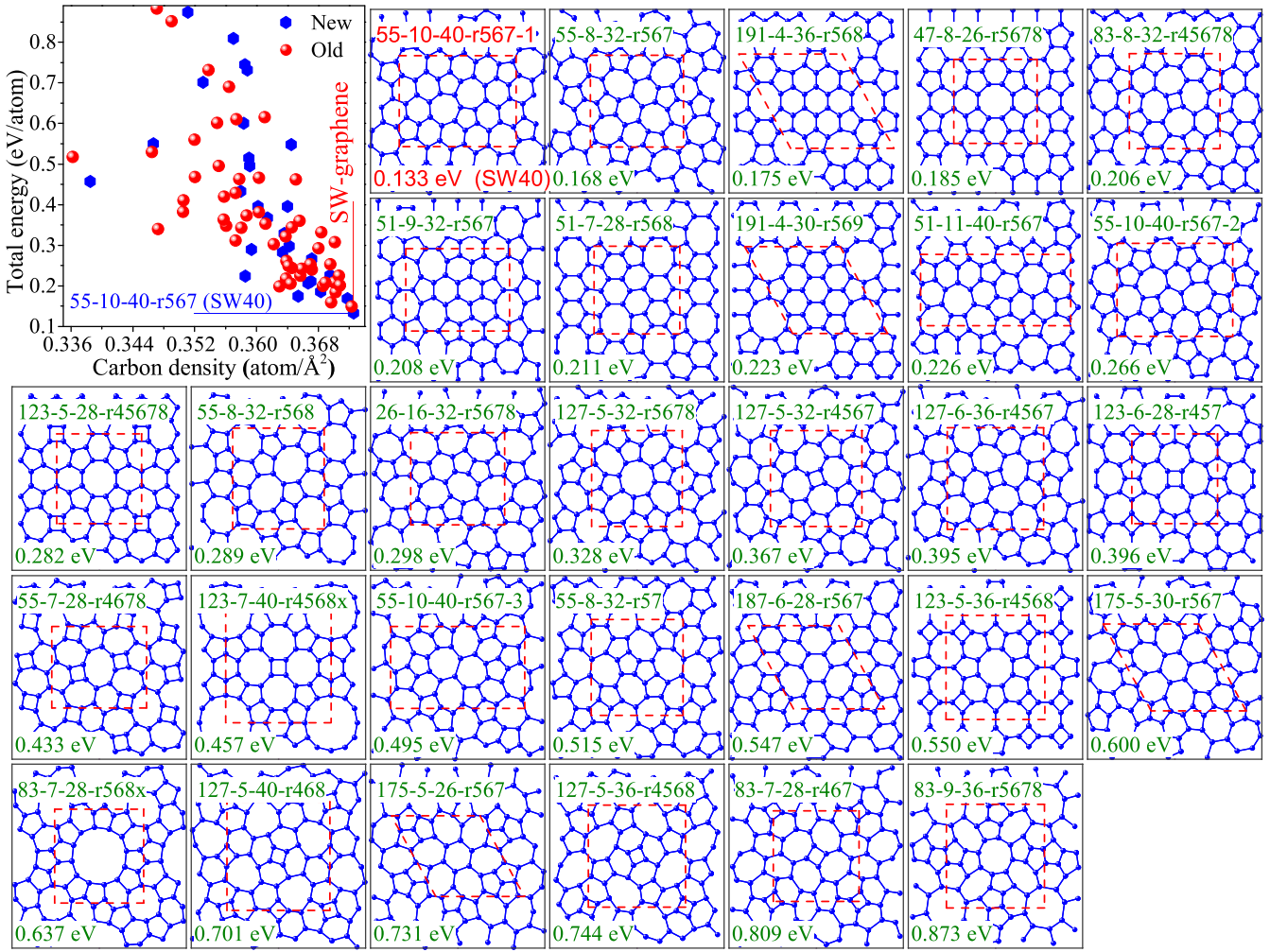}
\caption{Part I: The calculated total energies of all the previously predicted (red solid circles) and the newly discovered (blue solid hexagons) 2D carbon allotropes plotted as functions of the corresponding carbon densities. Part II: the optimized crystalline structures of the 30 newly discovered 2D carbons together with their names (sn-in-tn-rs) and energies (eV per atom). The nomenclature of "sn-in-tn-rs" is designed for revealing the fundamental crystal information of the space group number (sn), inequivalent atom number (in), total atom number (tn) and carbon ring sequence (rs).}
\end{center}
\end{figure*}

In this work, we report our discover of a new Stone-Wales graphene (SW40) with intrinsic type-III Dirac-cone and excellent stability. SW40 contains 40 carbon atoms in its orthorhombic crystalline cell and well-arranged Stones-Wales patterns. Its calculated total energy is only 133 meV/atom higher than that of graphene and lower than those of all the previously predicted 2D carbons. The calculated electronic properties reveal that SW40 processes intrinsic type-III Dirac-cone formed by band-crossing of a local linear-band and a local flat-band. Such an intrinsic type-III Dirac-cone can be effectively tuned by the in-plane strains and it will be transferred into Type-II and Type-I Dirac-cones under tensile and compressed strains, respectively. Finally, a general tight-binding model was constructed to understand the Dirac-cone features in SW40, which can be understood by including the interactions between adjacent Stone-Wales defects.

To search new 2D carbon allotropes with both remarkable stability and highly anisotropic Dirac-cone, we performed stochastic group and graph constrained searches \cite{jpcm,prb2004}based on our previously developed RG$^2$ code \cite{yhcprb, Shi18,hcyprl}, which is a high-efficient code for generating crystal structures with well-defined structural-feature\cite{splpb,jnpss,lzqass,znprb,yxjap}. The first-principles calculations are performed by the widely used VASP code \cite{VASP}, with the projector augmented wave methods (PAW)\cite{PAW1,PAW2} and the generalized gradient approximation (GGA) \cite{PBE}. A plane wave basis with cutoff energy of 500 eV is used to expand the wave functions for all carbon systems and the Brillouin zone sampling meshes are set to be dense enough to ensure the convergence (11$\times$11$\times$1 for SW40). All the 2D carbon structures are fully optimized until the residual forces on every atom is less than 0.001 eV/${\AA}$. The convergence criteria of total energy is set to be 10$^{-7}$ eV and the thickness of the slab-model is set to be larger than 15 ${\AA}$ to avoid spurious interactions between adjacent images. The open-free PHONOPY code \cite{phonopy} is employed to simulate the vibrational spectrum of SW40.

\begin{figure*}
\begin{center}
\includegraphics[width=\textwidth]{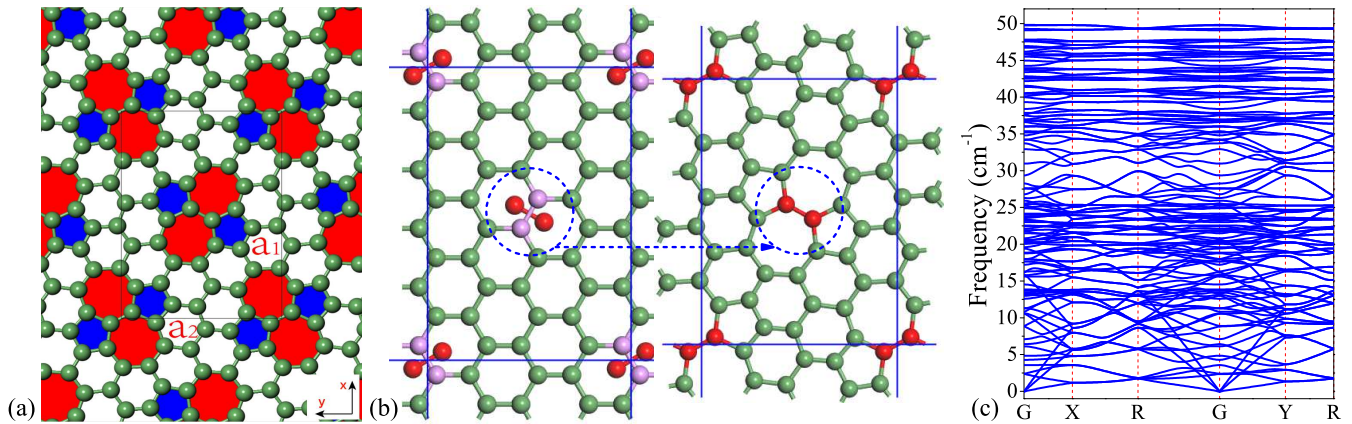}
\caption{(a) The optimized crystal structure of SW40 with well-arranged Stone-Wales defects as colored in blue and red. (b) The potential structural transition pathway from a given graphene supercell to SW40. (c) The calculated phonon band structure of SW40.}
\end{center}
\end{figure*}
Our RG$^2$ generates 90 unequal 2D sp$^2$ carbons including most of all the previously predicted ones and 30 new allotropes as shown in Fig. 1 and Fig. S1. To reveal their fundamental structural information, all the 2D carbon allotropes are named as "sn-in-tn-rs" according to their space group numbers (sn), inequivalent atom numbers (in), total atom numbers (tn) and ring sequences (rs). For the newly discovered 30 2D carbons, their names are shown in Fig.1 together with their optimized crystal structures and the corresponding total energies relative to graphene. We can see that these newly discovered 2D carbon allotropes are all three-connected and they contain mixed 4-, 5-, 6-, 7-, 8-, 9-member rings in their bodies. It is noticed that 55-10-40-r567-1 can be considered as a new Stone-Wales graphene \cite{yhcprb} with well-arranged Stone-Wales defects in its body as shown in Fig. 2 (a) and we further share it a short name of SW40 according to its total atom number. As shown in Fig. 2 (b), SW40 possesses another structural feature that it can be structurally constructed through a 90$^\circ$-rotation of carbon-dimmer in a given orthorhombic supercell of graphene. Such a structural feature is very similar to the previously predicted SW-graphene \cite{yhcprb}, Pza-C10\cite{yhc31}, R$_{57-2}$\cite{yhc17} and OPG-Z\cite{prb2013} as shown in Fig. S2.

To discuss the relative stabilities of these newly discovered 2D carbon allotropes, their calculated total energies relative to graphene are plotted in Fig .1 together with the corresponding carbon densities. The results show that most of these newly discovered 2D carbons are more stable than the previously proposed T-graphene (518 meV/atom) \cite{yhc26, yhc27, yhc28}. Especially, the total energy of SW40 is only about 133 meV higher than that of graphene and it is lower than those of all the previously predicted 2D carbon structures, including the recently synthesized phagraphene (201 meV/atom) and TPH-graphene (419 meV/atom) \cite{jacs}. Such an excellent stability indicate that SW40 is an expectable target for future experiment.

The vibrational spectrum of SW40 was calculated to evaluate its dynamical stability as shown in Fig. 2 (c). There is no any negative frequency appearing in phonon band structure, indicating that SW40 is dynamically stable under small vibration. The thermal stability of SW40 is also investigated by ab initio molecular dynamics simulations through a 2$\times$3$\times$1 supercell. After heating at 300 K and 500 K for 5 ps with a time step of 1 fs, the structure of SW40 remains intact and the total energies only oscillate about a constant value as shown in Fig. S3. Such results suggest that SW40 crystal is thermally stable and room temperature. Since 2D graphene with Stone$-$Wales defects have been synthesized in experiments \cite{PRL11} and the periodic non-hexagonal carbon-rings have been realized in graphene nanoribbons \cite{jacs, nc1}, SW40 is expected to be realized in future. It is noticed that the previously proposed phagraphene \cite{yhc36} and TPH-graphene \cite{jacs} was recently synthesized \cite{jacs} by the widely-used molecular assembly \cite{yhc46, yhc51} method. Thus, it is highly desirable that SW40 can be realized through assembly of ethylene and benzene (or naphthalene and azulene) as shown in Fig.~S4.

Fig. 3 (a) shows the DFT-based band structure of SW40 in the first BZ along high symmetric k-path $\Gamma(0,0)$ $\rightarrow$ $X(0.5,0)$ $\rightarrow$ $R(0.5,0.5)$ $\rightarrow$ $Y(0,0.5)$. It is interesting that there is a distorted Dirac-point (locating at D $(0, 0.115)$) formed by the crossing of a local linear-band and a local flat-band at the Fermi-level. To further confirm such a band-crossing, the decomposed charge density nearby the Dirac-point are calculated and shown in Fig. S5. One can see that the charge densities at the two sides of the Dirac-point are inverted, suggesting that the Dirac-point in SW40 is formed by band-crossing. Furthermore, the calculated density of state (DOS) at the Fermi level with value of zero can also confirm that SW40 is a semimetal with Dirac-point formed by band-crossing.

The DFT-based 3D surface band structure of SW40 nearby the Dirac-point is plotted in Fig. 3 (b) and it clearly shows that there is a distorted Dirac-cone. We can see that the distorted Dirac-cone in SW40 is formed by the crossing of a local linear-band and a local flat-band, which is some different from the Type-I cones in graphene, phagraphene and SW-graphene. As discussed by Liu et, al. in their recent work \cite{PRL2018Liu}, such a distorted Dirac-cone with local flat band can be classified as type-III Dirac-cone. And it can be realized in compressively strained BP under a periodic field of a circularly polarized laser to provide us new type of Fermion quasiparticles. We notice that such a new type of Dirac-cone (type-III) in the SW40 is intrinsic and it had been reported in previously proposed OPG-Z \cite{prb2013}. Given its excellent stability exceeding than OPG-Z (348 meV/atom), SW40 (133 meV/atom) is highly expected to be an experimental target with intrinsic type-III Dirac-cone for realizing such new Fermion quasiparticles.

Another significant feature of the type-III Dirac-cone is the high anisotropy. As shown in the 3D surface band structure, both the occupied and unoccupied bands nearby the Dirac-point show very different slope in different directions. To evaluate the anisotropy of the type-III Dirac-cone in SW40, we calculate the direction-dependent Fermi velocities in the cone as ${v_{f}=E(k)/\hbar|k|}$. As show in Fig. 3 (c), the Fermi velocity of the new Fermions in SW40 nearby the Dirac-Cone is strongly anisotropic with a heart-like shape. They are characterized by a strongly varying Fermi velocities from $2.611\times10^3m/s$ along the ${-}y ({\theta=90^{\circ}})$ direction to $5.697\times10^5m/s$ along to the ${+}y ({\theta=270^{\circ}})$ direction. The anisotropy of the Dirac-cone (A(D)) can be defined as\cite{EPL14}: A(D)=(v$_{max}$($\theta$)-v$_{min}$($\theta$))/(v$_{max}$($\theta$)+v$_{min}$($\theta$)), where v$_{max}$($\theta$) and v$_{min}$($\theta$) are the maximum and minimum of the direction-dependent Fermi velocities in the cone. Based on the data in Fig. 3 (c), we can easily obtain the anisotropy of the type-III Dirac-cone in SW40 of A(D) =$99.09\%$, which is gigantic in comparing with those in graphene and graphyne type materials \cite{EPL14}. The intrinsic type-III Dirac-cone with strong anisotropy, free from extra manipulating efforts, will broaden its potential applications in information applications.
\begin{figure}
\begin{center}
\includegraphics[width=\columnwidth]{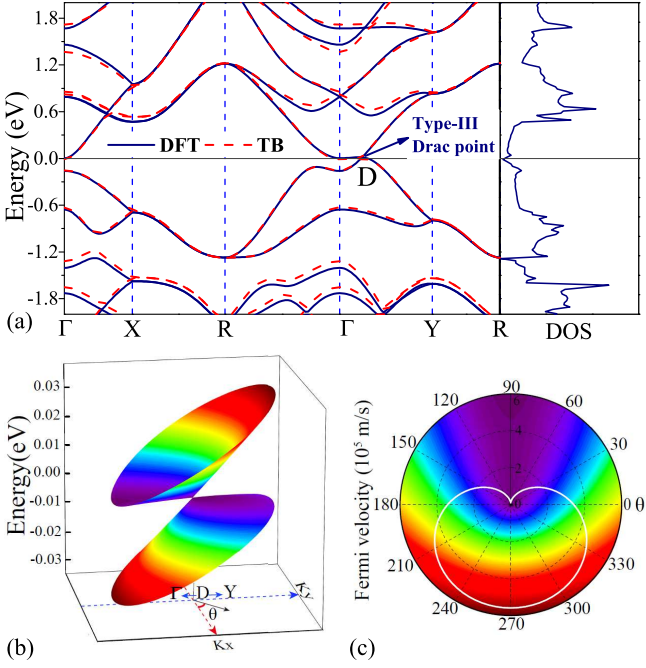}
\caption{(a) Electronic band structures and Density of states of SW40 from DFT (blue) and TB (red) calculations. (b) The 3D band structures of SW40 nearby the Dirac-point to show its Dirac-cone feature. (c) The direction-dependent Fermi velocities calculated based on DFT results (white line).}
\end{center}
\end{figure}
\begin{figure}[h]
\begin{center}
\includegraphics[width=\columnwidth]{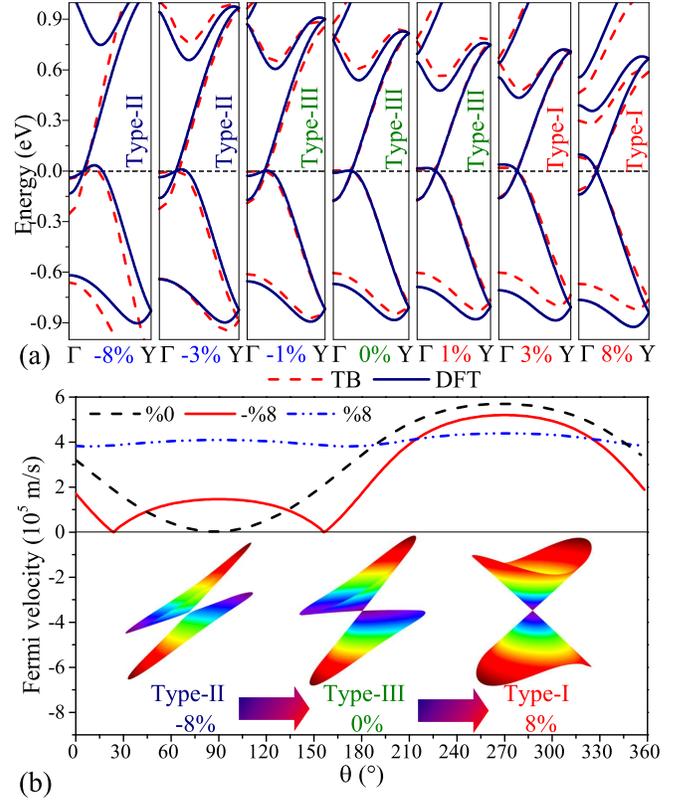}
\caption{(a) The DFT-based (blue solid lines) and TB-based (red dash lines) band structures of SW40 under selected strains (-8$\%$, -3$\%$, -1$\%$, 0$\%$, 1$\%$, 3$\%$ and 8$\%$). (b) The direction-dependent Fermi velocities in type-II (-8$\%$), type-III (0$\%$) and type-I (8$\%$) Dirac-cones and the corresponding 3D plot of the cones.}
\end{center}
\end{figure}

It is interesting that the properties of the type-III Dirac-cone in SW40 can be effectively tuned by external strains. As shown in Fig. 4 (a) are partial band structures of SW40 along $\Gamma$-Y k-path under selected biaxial in-plane strains (-8$\%$, -3$\%$, -1$\%$, 0$\%$, 1$\%$, 3$\%$ and 8$\%$). We can see that the Dirac-cone in SW40 can survive in different strains but the curvatures of the two energy bands nearby the crossing-point are significantly changed, resulting in two transitions from type-III to type-I in tensile strains and type-III to type-II in compressive strains, respectively. Under tensile strains, it is found that the curvatures of the two bands nearby the crossing-point are just slightly modulated. However, these two crossing bands move towards to each other (see from Y point) and correspondingly shift the crossing-point far away from the local flat-band to gradually form a type-I Dirac-cone under tensile strains.

The shapes of the two crossing bands are obviously changed and they move depart from each other (see from the Y point) under compressive stains, which destroy the type-III Dirac-cone feature and form a new type-II Dirac-point in $\Gamma$-Y. As shown in Fig. 4(b), we have also calculated 3D band structures to confirm the Dirac-cone features in SW40 under strains -8$\%$, 0$\%$ and 8$\%$, respectively. The corresponding direction-dependent Fermi velocities in Fig. 4 (b) show that tensile strains will weaken the anisotropy of the Dirac-cone and the compressive strains will enhance such a anisotropy. In Fig. S6, the results suggest that the type-III Dirac-cone feature in SW40 can survive in the strains range from -1$\%$ to 2$\%$, the transition points from type-III to type-II and type-I are compressive -2$\%$ and tensile 3$\%$, respectively.
To further understand the DFT-based results, we adopted a general tight-binding (TB) model based on the Slater-Koster method to describe the electrons in SW40 with energy nearby the Fermi-level. As the electronic states around the Fermi-level are predominantly contributed by p$_z$ orbitals of C atom, only C-p$_z$ orbital is included in the TB model. The TB Hamiltonian could be generally written as
\begin{equation}
 \begin{split}
&H=-\sum_{i\neq j}t_{ij}(c_{i}^{\dagger}c_{j}+h.c.)+\varepsilon_{\pi}\sum_{i}c_{i}^{\dagger}c_{i} \\
&S~=~~\sum_{i\neq j}s_{ij}(c_{i}^{\dagger}c_{j}+h.c.)+\sum_{i}c_{i}^{\dagger}c_{i}
 \end{split}
\end{equation}
where $t_{ij}$ and $s_{ij}$ are the hopping integrals and overlap integrals of an electron between the $i$-th and $j$-th atoms, respectively; $c_{i}^{\dagger}$ and $c_{j}$ are the creation and annihilation operators. Because all the atoms are carbons, the on-site energy $\varepsilon_{\pi}$ can be set to 0 eV. The distance-dependent hopping integrals and overlap integrals are determined by the formula
\begin{equation}
\begin{split}
t_{ij}=V_{pp\pi}e^{q_{1}\times(1-d_{ij}/d_{0})}; s_{ij}=S_{pp\pi}e^{q_{2}\times(1-d_{ij}/d_{0})}
\end{split}
\end{equation}
$d_{0}$ is the reference distance and set to be the length of standard C-C bond (1.426 \AA) in graphene, $V_{pp\pi}$ and $S_{pp\pi}$ are the hopping integrals and overlap integrals between p orbitals at $d_0$. $d_{ij}$ is the distance between the $i$-th and $j$-th atom and $q_{1,2}$ are the decay constants for the integrals. As $d_{ij}$ is larger than the used cutoff-distance $d_{cut}$, these integrals are set to be 0. These tight{-}binding parameters are fitted by minimizing the standard deviation function $\chi(V_{pp\pi},S_{pp\pi},q_{1},q_{2})$ as:
\begin{equation}
 \begin{split}
 \chi(\sigma)=\sqrt{\frac{1}{N_{data}}\sum_{n,k}[\mu_{n,k}\times(E_{n,k}^{TB}-E_{n,k}^{DFT})]^2}
 \end{split}
\end{equation}
where $E_{n,k}^{TB}$ and $E_{n,k}^{DFT}$ are the n-th TB-based and DFT-based eigenvalues relative to the corresponding Fermi-level at the used k-point. $\mu_{n,k}$=1 ($|E_{n,k}^{DFT}|$ $<$ $\sigma$) or 0 ($|E_{n,k}^{DFT}|$ $>$ $\sigma$), depending on the used cutoff energy $\sigma$. $N_{data}$ is the total number of data points in the energy area from -$\sigma$ eV to $\sigma$ eV for all the used k-points.

We have fit the above parameters with the cutoff-distance (d$_{cut}$) range from 2 {\AA}, 4 {\AA}, 8 {\AA}, 10 {\AA} and 12 {\AA} (as indicated in Fig.S7) by minimizing the deviation function x(vc) ($\sigma$$\approx$1.2) of the first valance band and the first conduction band. The finally optimal parameters under d$_{cut}$s are summarized in Tab.I together with their $\chi$(vc) and the corresponding band structures are shown in Fig. S8 (a)-(h) compared with the DFT-based results. We can see that under d$_{cut}$ of 2 {\AA} the optimized parameters can not describe the band structures of SW40 well ($\chi$(vc)=88 meV, high-anisotropy, type-I). The optimized parameters under d$_{cut}$s of 4 {\AA} and 6 {\AA} can approximatively describe the band structures nearby the Fermi-level with standard deviations $\chi$(vc) of about 38 meV and 27 meV, respectively. However, they both are type-I with high-anisotropy and can't fit the DFT-based type-III Dirac-point. Parameters obtained under d$_{cut}$s of 8 {\AA}, 10 {\AA} and 12 {\AA} can fit the DFT-results well, resulting in type-III Dirac-point and low standard deviation $\chi$($\sigma$) of about 18 meV. Such an excellent agrement can be understood according to the fact that these d$_{cut}$s are greater than 7.445 {\AA}, which are large enough to include the interactions between adjacent Stone-Wales defects.

The $\sigma$-dependent standard deviations $\chi$($\sigma$) shown in Fig. S7 suggest that the parameters optimized under d$_{cut}$s of 8 {\AA}, 10 {\AA} and 12 {\AA} can fit the DFT-based band structures very well in the energy range from -1.5 eV to 1.5 eV. In larger energy range from -3.75 eV to 3.75 eV, these parameters can also repeat the DFT-based band structure well with slightly enlarged standard deviations $\chi$($\sigma$). As shown in Fig. S9, the parameters obtained from d$_{cut}$ smaller than 7.445 {\AA} can not fit the DFT-based bands when we use larger d$_{cut}$s for calculation. The parameters can be improved at d$_{cut}$ of 8 {\AA} and the standard deviation $\chi$(vb) can achieve good convergence in 10 {\AA} and 12 {\AA}. For sp$^2$ hybridized carbon systems, we care about only the energy area nearby the Fermi-level, especially the first valance band and the first conduction band. Thus, we finally suggest using the parameters optimized under d$_{cut}$ of 10 {\AA} to investigate the band structure of SW40. Based on such optimal parameters under d$_{cut}$ of 10 {\AA}, we calculated the band structures of SW40 using different d$_{cut}$s of 8 {\AA}, 6 {\AA}, 5 {\AA}, 4 {\AA}, 3 {\AA} and 2 {\AA} as shown in Fig. S10 together with the corresponding standard deviations. We can see that it is difficult to describe the band structure of SW40, especially the type-III Dirac-cone feature, if we do not include the interactions between adjacent Stone-Wales defects (7.445 {\AA}).

We have also estimated the transferability of the finally used parameters obtained from d$_{cut}$= 10 {\AA} by comparing the DFT-based and TB-based band structures of some previously predicted 2D carbons as shown in Fig. S11. The used testing structures include metals (Octite-M1\cite{yhc22} and H$_{567}$ \cite{yhc17}), semi-metals(Phagraphene \cite{yhc36} and SW-graphene \cite{yhcprb}) and semiconductors (Octite SC \cite{yhc23} and $\Theta$-graphene \cite{YWC2018, ESM2019}) with different symmetries (Hexagonal, Tetragonal and orthorhombic). We can see that the parameters can commendably fit the DFT-based band structures nearby the Fermi-level, leaving a relatively small standard deviations ($\chi$). However, we find that these parameters can not directly repeat the strain effects on the band structure of SW40 well. Thus, we further introduce two quadratic polynomials \cite{prbgg} to include additional strain-induced decaying-functions for $V_{pp\pi}$ and $S_{pp\pi}$ as discussed in the supplementary file. As shown in Fig.4 (a) and Fig. S6, the modulating effects of external strains on the band structures SW40, including the transition of type-III to type-I and type-III to type-II, can be well described.
\begin{table}
\center
\caption{The optimal values of V$_{pp\pi}$, S$_{pp\pi}$, q$_1$ and q$_2$ determined under different cutoff-distances (d$_{cut}$) and the corresponding standard deviations $\chi$(vc). The corresponding band structures are shown in Fig. S8 and their corresponding Dirac-cone type (DT) are marked in this table.}
\begin{tabular}{c c c c c c c c c c c c}
\hline \hline
&d$_{cut}$ &V$_{pp\pi}$  &S$_{pp\pi}$ &q$_1$ &q$_2$ &$\chi$(vc) &BS &DT\\
\hline
&2  &-2.354 &0.0356  &5.003  &1.216    &0.08803  &Fig. S8(a) &Type-I\\
&4  &-1.909 &0.4628  &1.794  &2.136    &0.03854  &Fig. S8(b) &Type-I\\
&6  &-2.431 &0.3151  &1.427  &1.349    &0.02702  &Fig. S8(c) &Type-I\\
&8  &-1.819 &0.4621  &1.358  &1.205    &0.01837  &Fig. S8(d) &Type-III\\
&10 &-1.786 &0.4767  &1.274  &1.183    &0.01855  &Fig. S8(e) &Type-III\\
&12 &-1.753 &0.4913  &1.192  &1.161    &0.01724  &Fig. S8(f) &Type-III\\
\hline \hline
\end{tabular}
\label{tab1}
\end{table}

In conclusion, we propose a new 2D carbon candidate SW40 by employing our previously developed RG$^2$ code and the widely used VASP code. SW40 contains 40 carbon atoms in its orthorhombic lattice with Pbam symmetry, forming well-arranged Stone$-$Wales pattern. Our results show that the total energy of SW40 is about 133 meV/atom above that of graphene, indicating that it is more stable than all the previously predicted 2D carbons. The detailed analysis of electronic properties reveals that SW40 processes intrinsic type-III Dirac-cone with high anisotropy. Such a type-III dirac-cone can by effectively modulated, with two transitions from type-III to type-I in tensile strains and type-III to type-II in compressive strains, respectively. Finally, a general tight-binding model based on the Slater-Koster method was constructed and our TB results reveals that the high anisotropic type-III Dirac-cone in SW40 can be well understood by the interaction between adjacent Stone-Wales defects.

This work is supported by the National Natural Science Foundation of China (Grants No. 11974300, 11974299, 11704319 and 11874316), the Natural Science Foundation of Hunan Province, China (Grant No. 2016JJ3118 and 2019JJ50577), and the Program for Changjiang Scholars and Innovative Research Team in University (No. IRT13093).

$\#$ These authors contributed equally to this work.


\bibliographystyle{apsrev}

\begin{thebibliography}{99}
\bibitem{yhc1}K. S. Novoselov, A. K. Geim, S. V. Morozov, D. Jiang, Y. Zhang, S. V. Dubonos, I. V. Grigorieva, and A. A. Firsov, Science, 306, 666 (2004).
\bibitem{yhc10}K. S. Kim, Y. Zhao, H. Jang, S. Y. Lee, J. M. Kim, K. S. Kim, J. H. Ahn, P. Kim, J. Y. Choi and B. H. Hong, Nature, 457, 706 (2009).
\bibitem{yhc11}Z. Z. Sun, Z. Yan, J. Yao, E. Beitler, Y. Zhu, J. M. Tour, Nature, 468, 549 (2010).
\bibitem{yhc2}G. Li, Y.Li, H. Liu, Y. Guo, Y. Li, and D. Zhu, Chem. Commun., 46, 3256 (2010).
\bibitem{yhc14}R. Matsuoka, R. Sakamoto, K. Hoshiko, S. Sasaki, H. Masunaga, K. Nagashio, and H. Nishihara, J. Am. Chem. Soc., 139, 3145 (2017).
\bibitem{yhc5}Y. Zhang, Y. -W. Tan, H. L. Stormer, and P. Kim, Nature, 438, 201 (2005).
\bibitem{yhc6}K. I. Bolotin,  F. Ghahari, M. D. Shulman, H. L. Stormer, P. Kim, Nature, 462, 196 (2009).
\bibitem{yhc7}K. S. Novoselov, Z. Jiang, Y. Zhang, S. V. Morozov, H. L. Stormer, U. Zeitler, J. C. Maan, G. S. Boebinger, P. Kim, and A. K. Geim, Science, 315, 5817 (2007).
\bibitem{yhc8}K. S. Novoselov, E. McCann, S. V. Morozov, V. I. Fal'ko, M. I. Katsnelson, U. Zeitler, D. Jiang, F. Schedin , and A. K. Geim, Nat. Phys., 2, 177 (2006).
\bibitem{kkk}R. V. Noorden. Moving towards a graphene world. Nature 442, 228 (2006).
\bibitem{yhc15}V. H. Crespi, L. X. Benedict, M. L. Cohen, and S. G. Louie, Phys. Rev. B, 53, R13303 (1996).
\bibitem{yhc17}H. Terrones, M. Terrones, E. Hern\'andez, N. Grobert,J-C. Charlier, and P. M. Ajayan, Phys. Rev. Lett., 84, 1716 (2000).
\bibitem{yhc18}G. Cs\'anyi, C. J. Pickard, B. D. Simons, and R. J. Needs, Phys. Rev. B, 75, 085432 (2007).
\bibitem{yhc19}X. Y. Li, Q. Wang, and P. Jena, J. Phys. Chem. Lett.,8, 3234 (2017).
\bibitem{carbon2019} X. Wang, Z. H. Feng, J. Rong, Y. N. Zhang, Y. Zhong, J. Feng, X. H. Yu, and Z. L. Zhan, Carbon 142, 438 (2019).
\bibitem{yhc20}M. T. Lusk, and L. D. Carr, Phys. Rev. Lett., 100, 175503 (2008).
\bibitem{yhc21}M. T. Lusk, and L. D. Carr, Carbon, 47, 2226 (2009).
\bibitem{yhc22}D. J. Appelhans, L. D. Carr, and M. T. Lusk, New. J. Phys., 12, 125006 (2010).
\bibitem{yhc23}D. J. Appelhans, Z. B. Lin, and M. T. Lusk, Phys. Rev. B, 82, 073410 (2010).
\bibitem{yhc26} J. Nisar, X. Jiang, B. Pathak, J. J. Zhao, T. W. Kang, and R. Ahuja, Nanotechnology, 23, 385704 (2012).
\bibitem{yhc27} X. Q. Wang, H. D. Li, and J. T. Wang, Phys. Chem. Chem. Phys., 14, 11107 (2012).
\bibitem{yhc28} Y. Liu, G. Wang, Q. S. Huang, L. W. Guo, and X. L. Chen, Phys. Rev. Lett., 108, 225505 (2012).
\bibitem{yhcprb} H. C. Yin, X. Z. Shi, C. Y. He, M. Martinez-Canales, J. Li, C. J. Pickard, C. Tang, T. Ouyang, C. X. Zhang, and J. X. Zhang, Phys. Rev. B, 99, 041405(R) (2019).
\bibitem{CMSncy2019} C. X. Zhao, Y. Q. Yang, C. Y. Niu, J. Q. Wang, and Y. Jia. Comput. Mater. Sci., 160, 115 (2019).
\bibitem{yhc31}X. G. Luo, L. M. Liu, Z. P. hu, W. H. Wang, W. X. Song, F. F. Li, S. J. Zhao, H. Liu, H. T. Wang, and Y. J. Tian, J. Phys. Chem. Lett., 3, 3373 (2012).
\bibitem{YWC2018} W. C. Yi, W. Liu, J. Botana, J. Y. Liu and M. S. Miao, J. Mater. Chem. A, 6, 10348 (2018).
\bibitem{ESM2019} S. W. Wang, B. C. Yang, H. Y. Chen, and E. Ruckenstein. Energy Storage Mater., 16, 619 (2019) .
\bibitem{prb2012} B. G. Kim, and H. J. Choi, Phys. Rev. B, 86, 115435 (2012).
\bibitem{njp2013} H. Q. Huang, W. H. Duan, and Z. R. Liu, New J. Phys., 15, 023004 (2013).
\bibitem{NPGcm2019} W. Liu, L. Zhao, E. Zurek, J. Xia, Y. H. Zheng, H. Q. Lin, J. Y. Liu, and M. S. Miao, npj Comput Mater 5, 71 (2019).
\bibitem{NPGcm2018} M. Park, Y. Kim, and H. Lee, npj Comput Mater 4, 54 (2018).
\bibitem{yhc36} Z. H. Wang, X. F. Zhou, X. M. Zhang, Q. Zhu, H. F. Dong, M. W. Zhao, and A. R. Oganov, Nano Lett., 15, 6182 (2015).
\bibitem{prb2013} C. Su, h. Jiang, and J. Feng, Phys. Rev. B, 87, 075453 (2013).
\bibitem{magic1e} Y. Cao, V. Fatemi, S. Fang, K. Watanabe, T. Taniguchi, E. Kaxiras, and P. Jarillo-Herrero, Nature, 556, 43 (2018).
\bibitem{magic2e} Y. Cao, V. Fatemi, A. Demir, S. Fang, S. L. Tomarken, J. Y. Luo, J. D. Sanchez-Yamagishi, K. Watanabe, T. Taniguchi, E. Kaxiras, R. C. Ashoori, and P. Jarillo-Herrero, Nature, 556, 80 (2018).
\bibitem{magic3e} A. Kerelsky, L. J. McGilly, D. M. Kennes, L. Xian, M. Yankowitz, S. W. Chen, K. Watanabe, T. Taniguchi, J. Hone, C. Dean, A. Rubio, and A. N. Pasupathy, Nature 572, 95 (2019).
\bibitem{magic4e} M. Yankowitz, S. W. Chen, H. Polshyn, Y. X. Zhang, K. Watanabe, T. Taniguchi, D. Graf, A. F. Young, and C. R. Dean, Science, 363, 1059 (2019)
\bibitem{magic1t} H. C. Po, L. J. Zuo, A. Vishwanath, and T. Senthil, Phys. Rev. X, 8, 031089 (2018).
\bibitem{magic2t} C. C. Liu, L. D. Zhang, W. Q. Chen, and F. Yang, Phys. Rev. Lett., 121, 217001 (2018).
\bibitem{magic3t} F. C. Wu, A. H. MacDonald and I. Martin, Phys. Rev. Lett. 121, 257001 (2018).
\bibitem{CYPprb} Y. P. Chen, S. L. Xu, Y. Xie, C. Y. Zhong, C. J. Wu, and S. B. Zhang, Phys. Rev. B, 98, 035135 (2018).
\bibitem{NP08} C. H. Park, L. Yang, Y. W. Son, M. L. Cohen, and S. G. Louie, Nature Phys., 4. 213 (2008)
\bibitem{PRL08} C. H. Park, L. Yang, Y. W. Son, M. L. Cohen, and S. G. Louie, Phys. Rev. Lett., 101, 126804 (2008).
\bibitem{NC14} F. N. Xia, H. Wang, and Y. C. Jia, Nat. Commun., 5, 4458 (2014).
\bibitem{Gdoping} H. H. Zhang, Y. Xie, C. Y. Zhong, Z. W. Zhang, and Y. P. Chen, J. Phys. Chem. C, 121, 12476 (2017).
\bibitem{O17} N. V. R. Nulakani and V. Subramanian, ACS Omega, 2, 6822 (2017).
\bibitem{JPCL15} L. Z. Zhang, Z. F. Wang, Zhiming M. Wang, S. X. Du, H.-J. Gao, and Feng Liu, J. Phys. Chem. Lett., 6, 2959 (2015).
\bibitem{PRB2016II} H. Q. Huang, S. Y. Zhou and W. H. Duan, Phys. Rev. B, 94, 121117(R) (2016).
\bibitem{PRX2016} L. Muechler, A. Alexandradinata, T. neupert and R. Car, Phys. Rev. X, 6, 041069 (2016).
\bibitem{PRB2017} A. S. Cuamba, P. Hosur, H. Y. Lu, L. Hao and C. S. Ting, Phys. Rev. B, 96, 195159 (2017).
\bibitem{NP} K. Deng, G. L. Wan, P. Deng, K. N. Zhang, S. J. Ding, E. Y. Wang, M. Z. Yan, H. Q. Huang, H. Y. Zhang, Z. L. Xu, J. Denlinger, A. Fedorov, H. T. Yang, W. H. Duan, H. Yao, Y. Wu, S. S. Fan, H. J. Zhang, X. Chen and S. Y. Zhou, Nat. Phys., 12, 1105 (2016).
\bibitem{PRL2018Liu} H. Liu, J. T. Sun, C. Cheng, F. Liu, and S. Meng, Phys. Rev. Lett., 120, 237403 (2018).
\bibitem{jacs} Q. T. Fan, D. Martin-Jimenez, D. Ebeling, C. K. Krug, L. Brechmann, C. Kohlmeyer, G. Hilt, W. Hieringer, A. Schirmeisen and J. M. Gottfried, J. Am. Chem. Soc., 141, 17713 (2019).
\bibitem{jpcm} C. J. Pickard, and R. J. Needs,J. Phys.: Condensed Matter, 23,053201 (2011).
\bibitem{prb2004} R. T. Strong, C. J. Pickard, V. Milman, G. Thimm, B. Winkler, Phys. Rev. B, 70, 045101 (2004).
\bibitem{Shi18} X. Z. Shi, C. Y. He, C. J. Pickard, C. Tang, and J. X. Zhong, Phys. Rev. B, 97, 014104 (2018).
\bibitem{hcyprl} C. Y. He, X. Z. Shi, S. J. Clark, J. Li, C. J. Pickard, T. Ouyang, C. X. Zhang, C. Tang, and J. X. Zhong, Phys. Rev. Lett., 121, 175701 (2018).
\bibitem{splpb}P. L. Sun, C. Y. He, C. X. Zhang, H. P. Xiao and J. X. Zhong, Physica B: Condens. Matt., 562, 131 (2019).
\bibitem{jnpss}N. Jiao, P. Zhou, C. Y. He, J. J. He, X. L. Liu and L. Z. Sun, Phys. Status. Solidi-RRL, 13, 1900470 (2019).
\bibitem{lzqass}Z. Q. Li, X. Z. Shi, C. Y. He, T. Ouyang, J. Li, C. X. Zhang, S. F. Zhang, C. Tang, R. A. Ro\"mer and J. X. Zhong, Appl., Surf. Sci., 497, 143803 (2019).
\bibitem{znprb} N. Zhou, P. Zhou, J. Li, C. Y. He and J. X. Zhong, Phys. Rev. B, 100, 115425 (2019)
\bibitem{yxjap} X. Yang, C. Y. He, X. Z. Shi, J. Li, C. X. Zhang, C. Tang and J. X. Zhong, J. Appl. Phys., 124, 163107 (2018).
\bibitem{VASP}G. Kresse, and J. Furthm\"uller, Phys. Rev. B, 54, 11169 (1996).
\bibitem{PAW1}P. E. Bl\"ochl, Phys. Rev. B, 50, 17953 (1994).
\bibitem{PAW2}G. Kresse, and D. Joubert, Phys. Rev. B, 59, 1758 (1999).
\bibitem{PBE}J. P. Perdew, K. Burke, and M. Ernzerhof, Phys. Rev. Lett., 77, 3865 (1996).
\bibitem{phonopy} Phonopy: http://atztogo.github.io/phonopy/
\bibitem{PRL11} J. Kotakoski, A. V. Krasheninnikov, U. Kaiser, and J. C.Meyer, Phys. Rev. Lett. 106, 105505 (2011).
\bibitem{nc1} M. Z. Liu, M. X. Liu, L. M. She, Z. Q. Zha, J. L. Pan, S. C. Li, T. Li, Y. Y He, Z. Y. Cai, J. B. Wang, Y. Zheng, X. H. Qiu and D. Y. Zhong, Nat. Commun., 8, 14924 (2017).
\bibitem{yhc46}J. V. Barth, G. Costantini and K. Kern, Nature, 437, 671 (2005).
\bibitem{yhc51}L. Grill, M. Dyer, L. Lafferentz, M. Persson, M. V. Peters and S. Hecht, Nat. Nanotechnol. 2, 687 (2007).
\bibitem{EPL14} D. Z. Yang, M. S. Si, G. P. Zhang, and D. S. Xue, Europhys. Lett. 107, 20003 (2014).
\bibitem{prbgg} G. Gui, J. Li and J. X. Zhong, Phys. Rev. B, 78, 075435 (2008).
\end{thebibliography}

\end{document}